# An Efficient Regional Storm Surge Surrogate Model Training Strategy Under Evolving Landscape and Climate Scenarios

## Preprinted Draft

Ziyue Liu[a,b,*]; Mohammad Ahmadi Gharehtoragh[a]; Brenna Kari Losch[a]; David R. Johnson[a,c]

[a]Edwardson School of Industrial Engineering, Purdue University, West Lafayette, IN, USA.

[b]Department of Civil and Environmental Engineering, University of Maryland, College Park, MD, USA.

[c]Department of Political Science, Purdue University, West Lafayette, IN, USA.

[*]Corresponding author: ziyue20@terpmail.umd.edu

**Abstract:**

Coastal communities face significant risk from storm-induced coastal flooding, which causes substantial societal and economic losses worldwide. Machine learning techniques have increasingly been integrated into coastal hazard modeling, particularly for storm surge prediction, due to advances in computational capacity. However, incorporating multiple projected future climate and landscape scenarios requires extensive numerical simulations of synthetic storm suites over large geospatial domains, resulting in rapidly escalating computational costs. This study proposes a cost-effective training data reduction strategy for machine learning–based storm surge surrogate models that enables efficient incorporation of new future scenarios while minimizing computational burden. The proposed strategy reduces training data across three dimensions: grid points, input features, and storm suite size. Reducing the storm suite size for future scenario simulations is highly effective in guiding numerical simulations, yielding substantial reductions in simulation cost. The performance of surrogate models trained on reduced datasets was evaluated using different machine learning algorithms. Results demonstrate that the proposed reduction strategy is robust across different model types. When trained using 5,000 out of 80,000 grid points, 10 out of 12 input features, and 60 out of 90 storms, the total training dataset is reduced to approximately 5% of its original size. Despite this reduction, the trained model achieves a correlation coefficient of 0.94, comparable to models trained on the full dataset. In addition, storm selection methodologies are introduced to support efficient storm set expansion for future scenario analyses.

**Highlights**

- Proposes a cost-effective data reduction strategy for storm surge ML surrogate models
- Reduces training data across grid points, features, and storm suite dimensions
- Storm suite reduction yields major savings in numerical simulation cost
- Comparable accuracy achieved with only ~5% of full training data (R = 0.94)
- Reduction strategy performs consistently across different ML algorithms

## 1. Introduction

Storm-induced coastal floods cause significant damage to coastal communities, infrastructure, and ecosystems. As a result, substantial efforts have been dedicated to assessing and managing these hazards. In recent decades, rapid advancements in computational power have facilitated the widespread adoption of machine learning (ML) techniques in coastal hazard modeling—particularly for storm surge prediction. Louisiana experienced severe impacts from Hurricane Katrina in 2005, which resulted in billions of dollars in damages. In response, the state established the Coastal Protection and Restoration Authority (CPRA). CPRA has initiated a series of long-term resilience efforts, including the development of the Louisiana Coastal Master Plan (CMP) (https://coastal.la.gov/our-plan/).

The CMP is revised on a six-year planning cycle (i.e., CMP2017, CMP2023, etc.) and consists of an approximately $50 billion US portfolio of recommended investments in coastal risk reduction and restoration projects to be implemented over the next 50 years. Each planning cycle has advanced new methods for estimating the hazard (i.e., annual exceedance probability distributions) associated with tropical cyclone impacts such as storm surge elevations, significant wave heights, and inundation depths (e.g., Johnson et al. 2013; Fischbach et al. 2016; Nadal-Caraballo et al. 2022; Gharehtoragh and Johnson 2024). As Louisiana's coastal regions continue to be exposed to land subsidence, and the loss of land and vegetation due to sea level rise, its landscape and regional natural systems remain highly sensitive to varying climate conditions and to local restoration efforts. At the same time, risk reduction measures (e.g., levees, floodwalls, pumps) are designed for multi-decadal useful lifetimes; it is therefore important to develop storm hazard estimates capable of accounting for varying climate and landscape futures in order to effectively design these projects.

A procedure for constructing storm surge hazard curves under evolving landscape and climate scenarios using storm surge surrogate models can consist of the key steps illustrated in Figure 1. In this procedure, a series of scenarios is designed to represent current and different projected future climate and landscape conditions. Numerical storm surge simulations are conducted under each scenario using a synthetic storm suite developed under the Coastal Hazard System–Louisiana (CHS-LA) study (Nadal-Caraballo et al. 2022). The resulting simulation outputs serve as training data for the development of a storm surge surrogate model. This storm surge surrogate model is designed to efficiently estimate storm surge responses across various climate and landscape scenarios. In the final step, the storm surge hazard curve is quantified using the developed surrogate models. Generally, two methods can be used for hazard curve quantification: (1) running the storm surge surrogate model with an augmented storm suite and constructing the hazard curve through the Joint Probability Method (JPM) integral (e.g., Liu et al. 2024b; Nadal-Caraballo et al. 2022); and (2) training a surrogate model that includes the return period as an input feature, enabling direct prediction of surge levels corresponding to different return periods. The training strategies for surrogate models applied in methods (1) and (2) differ slightly. In this study, we focus on the type of storm surge surrogate models employed in the first method. In the whole procedure, numerical simulation is the most computationally expensive step. While surrogate model training is relatively less expensive, it can still be time-consuming, particularly when the training dataset is large (Gharehtoragh and Johnson 2024).

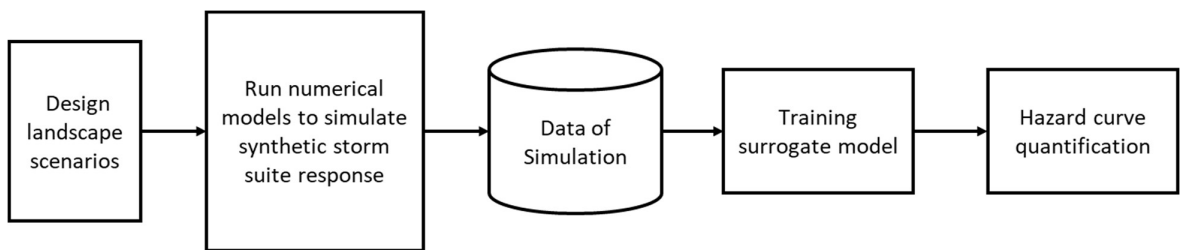

Figure 1: Key steps of hazard curve quantification for different scenarios.

The development of storm surge surrogate models is a well-established topic in the field of coastal hazard analysis (e.g., Al Kajbaf and Bensi 2020; Jia and Taflanidis 2013; Liu et al. 2024). To accelerate the training of regional storm surge surrogate models, various dimensionality reduction techniques have been explored and implemented (Jia et al. 2016; Jia and Taflanidis 2013). Principal Components Analysis (PCA) has been widely used to convert high-dimensional output, which is usually the storm surge response at each grid point (GP) of a mesh-gridded domain, into a latent space, thereby reducing computational cost. Building upon the PCA method, clustering methods have also been employed, either to improve prediction accuracy (Lee et al. 2021) or to enable interpolation for faster construction of regional hazard curves (Kyprioti et al. 2021). In this study, because the landscape features at each GP are included as input variables to predict the storm surge response at each individual GP (one-dimensional output) to enable predictions under varying landscape scenarios, the PCA method cannot be directly applied to downscale output dimensions. However, the PCA method is incorporated to support a GP reduction analysis as described in Section 3.1.

The objective of this study is to develop an efficient strategy for training storm surge surrogate models that is capable of predicting storm surge under varying future climate and landscape conditions. The proposed training framework is designed to be flexible and extensible, allowing for the continuous integration of additional training data from newly designed scenarios.

## 2. Method
### 2.1. Reference model and reduction strategy

Gharehtoragh and Johnson (2024) developed a storm surge surrogate model capable of estimating storm surge responses under varying climate and landscape scenarios for CMP2023. This model is treated as a reference model for this study. The reference model was trained using numerical simulation data from one base scenario representing the 2020 landscape and ten projected future scenarios with synthetic storm suites. The future scenarios represent decadal time slices from 2030 to 2070 under two different scenarios that adopt differing assumptions about environmental factors, such as the rate of sea level rise and land subsidence (CPRA 2023). The surrogate model employs a feed-forward artificial neural network (NN) architecture designed to incorporate landscape features at individual GPs. It adopts a one GP input to one GP output structure, allowing each prediction to reflect GP-specific conditions. The NN consists of four hidden layers, each containing 256 neurons. All hidden layers use the ReLU activation function, while the output layer uses a linear activation function to predict peak surge values. The model was trained using a learning rate of 0.001.

The input feature vector comprises 12 variables:

- Five storm parameters: central pressure ($P_c$), forward velocity ($V_f$), radius of maximum wind ($R_{max}$), landfall angle ($\theta$), and landfall longitude;
- Six GP-specific spatial and landscape parameters: latitude, longitude, Manning's $n$, canopy coefficient, surface roughness coefficient ($Z_0$), and topographic/bathymetric elevation;
- One climate condition parameter: mean sea level (MSL).

The model's output is the peak storm surge at each individual GP associated with a given synthetic storm in a given landscape scenario, enabling high-resolution surge estimation across varying scenarios.

In this study, building upon the reference surrogate model developed by Gharehtoragh and Johnson (2024), a reduction strategy for efficient training is proposed to construct a flexible training framework with enhanced computational efficiency. This reduction strategy is designed to support the continuous integration of newly generated scenario data. The reduction strategy involves sequentially applying three reduction approaches using available scenario datasets, followed by performance evaluation using newly generated scenario data. The key steps of this reduction strategy are illustrated in Figure 2. The three reduction approaches, which form the core components of the efficient training strategy, are as follows:

- Grid points reduction (Box 1): The reference model includes over 80,000 GPs taken from the Coastal Louisiana Risk Assessment (CLARA) model. The GPs form a mixed-resolution mesh with a minimum resolution of one km², with added resolution in populated areas such that every US census block contains at least one GP. To reduce computational burden while maintaining regional representativeness, clustering methods are employed to identify a representative subset of GPs across the study domain.
- Input features reduction (Box 2): The reference model utilizes 12 input variables, including storm parameters and landscape features. To reduce redundancy and improve training efficiency, a correlation analysis is conducted to select a simplified yet informative subset of input features.

- Storm set reduction (Box 3): Clustering and adaptive sampling techniques are explored to minimize the number of storms required for training, while preserving model accuracy and ensuring adequate diversity in storm characteristics.

Among these three reduction approaches, the storm reduction approach is particularly interesting, as it can directly guide the selection of storms for numerical simulation and thus substantially reduces computational costs. An evaluation step (Box 4) is incorporated into the framework to assess the performance of the efficient training strategy using newly generated scenario data.

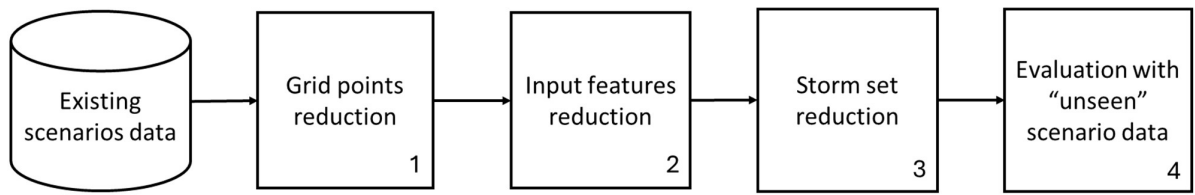

Figure 2: Key steps of reduction strategy.

## 2.2. Data used

In this work, a series of scenario datasets is utilized to evaluate storm surge responses under varying sea level rise conditions. These datasets were generated by running the ADCIRC model using synthetic storm suites for each projected landscape and sea level scenario. Table 1 summarizes the key information for each scenario dataset. The scenario IDs follow the format of S##Y##, where S## denotes the assumed sea level rise rate and Y## indicates the projection year. For example, S00 corresponds to a static, present-day sea level condition (i.e., MSL remaining at its 2020 value), whereas S09 represents the most extreme sea level rise assumption. Likewise, Y10 and Y50 correspond to projections for the years 2030 and 2070, respectively.

Among these scenario datasets, S00Y00 and S09Y50 include storm surge responses for the full synthetic storm suite of 645 storms developed in the CHS-LA study (Nadal-Caraballo et al. 2022). The other scenarios include a representative subset of 90 storms—referred to as the CMP2023 90 storm set—selected from the full 645-storm suite using an optimization algorithm designed to minimize errors in hazard curve integration during CMP2023 (Fischbach et al. 2021).

Table 1: Information of scenario datasets

| Scenario ID | Year | MSL (NAVD 88, m) | Number of Storms |
|---|---|---|---|
| S00Y00 | 2020 | 0.36 | 645 |
| S07Y10 | 2030 | 0.44 | 90 |
| S07Y20 | 2040 | 0.52 | 90 |
| S07Y30 | 2050 | 0.62 | 90 |
| S07Y40 | 2060 | 0.73 | 90 |
| S07Y50 | 2070 | 0.86 | 90 |
| S08Y10 | 2030 | 0.46 | 90 |
| S08Y20 | 2040 | 0.58 | 90 |
| S08Y30 | 2050 | 0.73 | 90 |
| S08Y40 | 2060 | 0.92 | 90 |
| S08Y50 | 2070 | 1.13 | 90 |
| S09Y50 | 2070 | 1.45 | 645 |

## 3. Reduction analysis

In this section, the reduction analysis is described in detail. The S00Y00, S07, and S08 scenario datasets are used in the reduction training process. As illustrated in Figure 2, the reduction procedure consists of three sequential steps: (1) GPs reduction (described in Section 3.1), (2) Input features reduction (Section 3.2), and (3) Storm set reduction (described in Section 3.3). It should be noted that the S08Y50 dataset is treated as an "unseen" scenario data during the reduction training process and is used exclusively in Section 3.4 for evaluating the trained model.

### 3.1. Grid points reduction

A regional storm surge surrogate model is typically developed for the purpose of predicting peak storm surge across a mesh-gridded domain, where its output dimensionality equals the number of GPs. PCA (Jia et al. 2016; Jia and Taflanidis 2013) has been commonly employed to reduce this high-dimensional output space to a lower-dimensional latent space. However, in this study, the storm surge surrogate model uses geospatial and landscape features of each individual GP as input and predicts a one-dimensional output (i.e., peak surge) per GP. As a result, PCA cannot be directly applied. Building upon the work of Kyprioti et al. (2021), a *k*-means clustering method is employed here to identify representative subsets of GPs for training storm surge surrogate models, aiming to reduce computational demand in the model training process. The input features for *k*-means clustering include geospatial features (latitude, longitude, and elevation), landscape features (canopy, Manning's *n* and $Z_0$), and the surge response-derived PCA eigenvector[1]. Because PCA requires a large and statistically diverse dataset to produce statistically meaningful results, the surge response PCA features are used exclusively for the *k*-means clustering of the S00Y00 scenario dataset, which includes simulations of 645 synthetic storms. For the S07 and S08 scenarios, only geospatial and landscape features are utilized for *k*-means clustering. Before conducting the *k*-means clustering, missing surge values at dry nodes for each scenario dataset are corrected using a *k*-nearest neighbors model with inverse distance weighting, following the method proposed by Jia et al. (2016). The subset of GPs is extracted by selecting the points closest to the centroid of each cluster. Figure 3 shows an example map of the clustered GPs and extracted points.

A sensitivity analysis is conducted to assess model performance under varying numbers of training GPs. A holdout cross-validation strategy is employed, where the model is trained on scenarios S00 Y00, S07Y10–Y50, and S08Y10–Y30 with reduced GPs, and tested on S08Y40 with the full 80,000 GPs. A leave-one-out cross-validation test, as conducted in Gharehtoragh and Johnson (2024), could serve as an alternative testing strategy to the holdout cross-validation strategy employed in this study if additional computational resources are available or if higher accuracy is needed. Two approaches for setting the GP subsets in different scenario data are compared: (1) Using a fixed subset of reduced GPs based on the *k*-means clustering result from the S00Y00 scenario dataset; (2) Using flexible subsets of reduced GPs extracted by running *k*-means clustering individually for each scenario dataset (i.e., extracting different GPs from each scenario dataset as training data). Figure 4 presents model performance under varying numbers of training GPs. It is found that when the number of training GPs is relatively small, the flexible centroid approach achieves superior performance. However, as the number of GPs increases, the performance difference between the two approaches becomes negligible.

---

[1] Consistent with the method introduced by Kyprioti et al. (2021), the eigenvalues are directly utilized to provide the prioritization for the different eigenvectors.

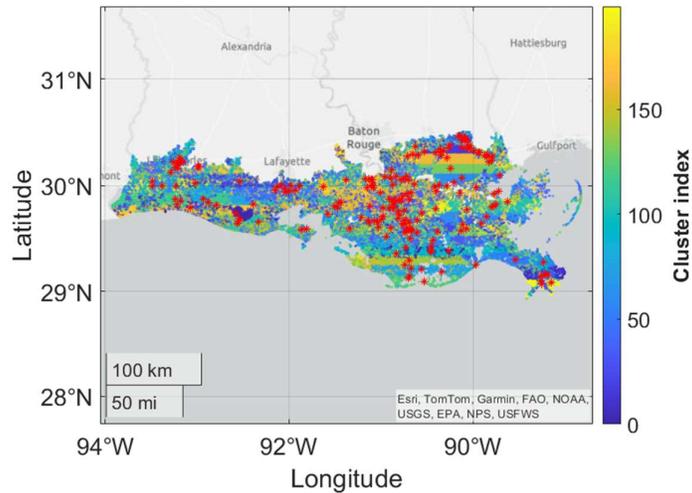

**Figure 3: Example map showing clustered GPs and extracted points.**

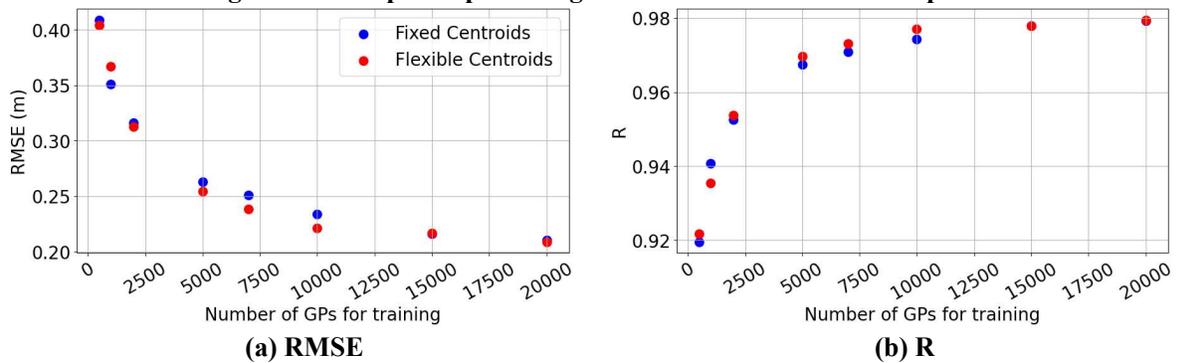

(a) RMSE
(b) R

**Figure 4: Model performance with different numbers of GPs involved for training.**

Different approaches involving inconsistent training GPs subset sizes across scenarios were explored in training. The motivation for using inconsistent subset sizes was based on the assumption that allocating more GPs to certain key scenarios—such as using 80,000 GPs in the S00Y00 scenario (which contains a larger storm suite) and in S08Y30 (which reflects a relatively high MSL condition within existing scenario datasets), while using only 5,000 GPs in the remaining scenarios—might enhance average model performance compared with consistently using 5,000 GPs for every scenario. However, the results indicate that employing inconsistent GP subset sizes across scenarios does not improve overall model accuracy. On the contrary, it introduces elevated prediction errors in certain sensitive regions. A comparison of spatial error maps is presented in Figure 5. As Figure 5 indicates, in the region along the Mississippi River west of New Orleans (circled), the error significantly increased when using an inconsistent training GP size. The observed degradation in performance can be attributed to the imbalance in training data caused by inconsistent GP allocation across different scenarios—an issue known to adversely impact regression models (Kowatsch et al. 2024).

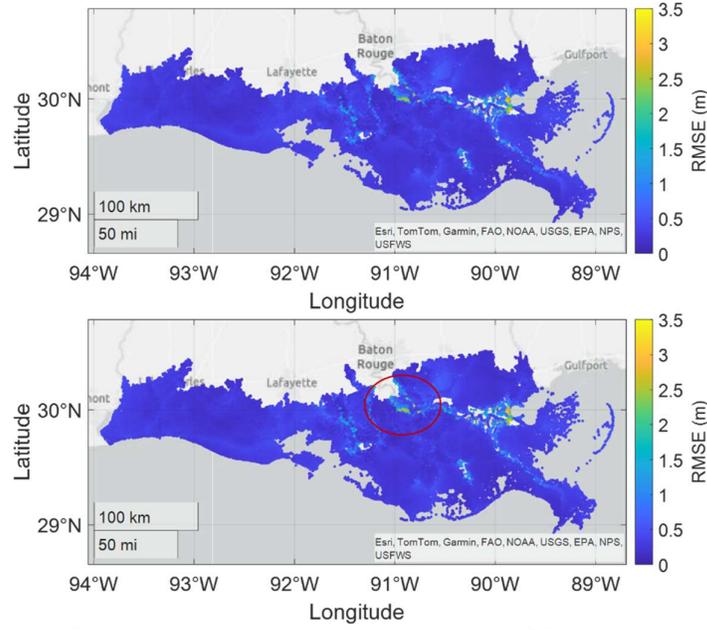

**Figure 5: Comparison of error maps using consistent training GP size and inconsistent training GP size. The red circle indicated a subregion with increased error when using an inconsistent training GP size.**

### 3.2. Input features reduction

The reference model incorporates 12 input features, which includes storm parameters ($P_c$, $V_f$, $R_{max}$, $\Theta$, and landfall longitude); GP spatial coordinates (latitude and longitude); landscape parameters Manning's $n$, canopy coefficient, $Z_0$, and elevation; and MSL as a global boundary condition. The storm parameter features—$P_c$, $V_f$, $R_{max}$, $\Theta$, and landfall longitude—have been widely used in surrogate model training for storm surge prediction (Al Kajbaf and Bensi, 2020). Latitude and longitude capture the geospatial location of each GP, while the landscape and climate parameters (canopy coefficient, Manning's $n$, $Z_0$, elevation, and MSL) describe environmental characteristics.

To investigate inter-feature relationships, a correlation analysis was conducted using all existing scenarios datasets with a fixed 5,000 GPs subset. The correlation matrix is presented in Figure 6. Strong positive and negative correlations are observed among the canopy coefficient, $Z_0$, and Manning's $n$. This is expected, as all three variables describe local vegetation characteristics, which are defined in a scenario-dependent manner. Figure 7 plots the pattern of these three variables to visualize the similarity among them. A moderate correlation is observed between $P_c$ and $R_{max}$, consistent with established tropical cyclone (TC) parameter relationships (Vickery and Wadhera 2008). Weak correlations are found between latitude and the landscape parameters, likely reflecting latitudinal variations in vegetation type and ground elevation. It is also noteworthy that MSL shows weak correlations with $P_c$ and $R_{max}$. This arises because the MSL values in the S07 and S08 scenarios are higher than those in the S00Y00 scenario dataset, and in the CPM2023 90 storm subset (storm set of the S07 and S08 scenario datasets), the proportion of storms with relatively high $P_c$ and low $R_{max}$ is greater than in the full 645 storm set (storm set of the S00Y00 scenario dataset).

To assess the potential effect of multicollinearity on model performance, ablation tests were performed by removing the canopy coefficient and $Z_0$ from the input features. As shown in Figure 8, the model's R and RMSE exhibited negligible changes when the canopy coefficient and $Z_0$ are removed. This result aligns with existing studies (e.g., Dormann et al. 2013) that discuss the potential redundancy caused by including highly correlated predictors in regression-based ML models.

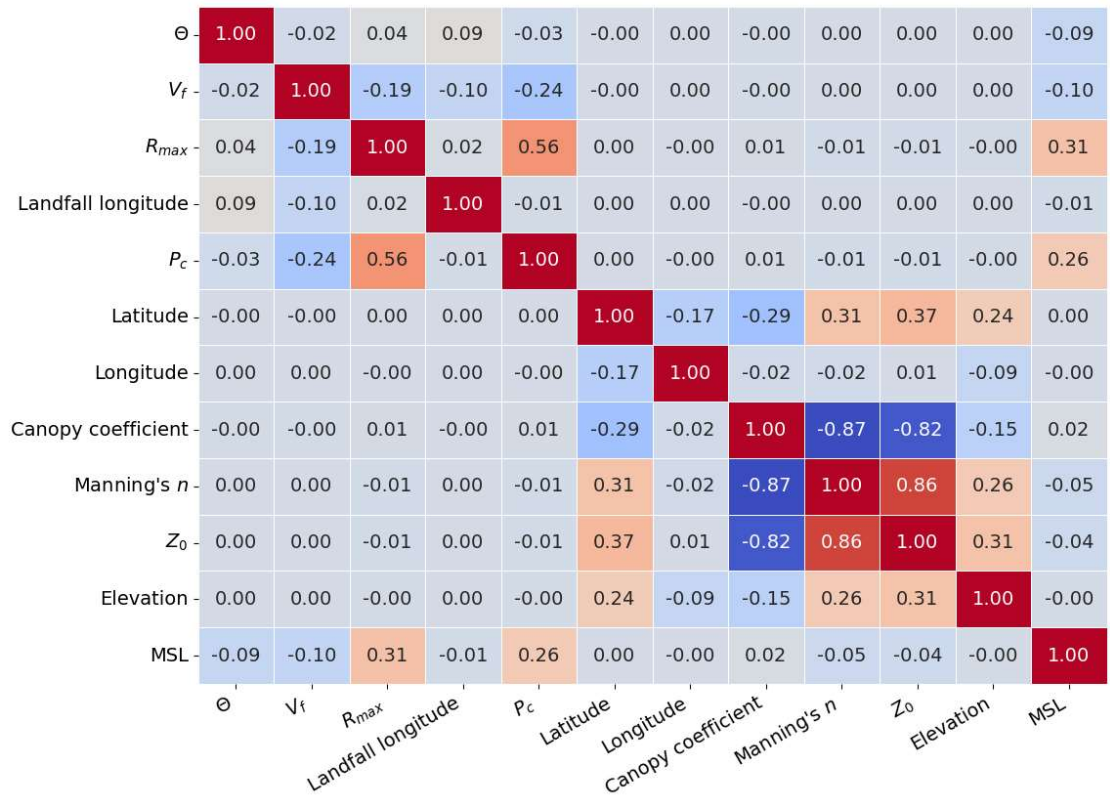

**Figure 6: Correlation matrix of input features.**

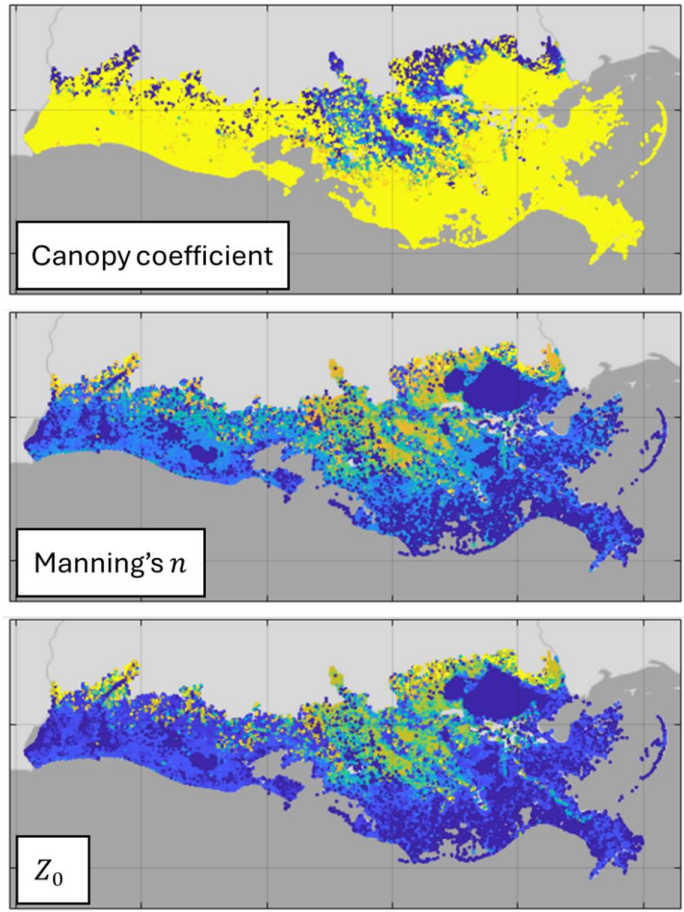

Figure 7: Geospatial pattern of landscape features.

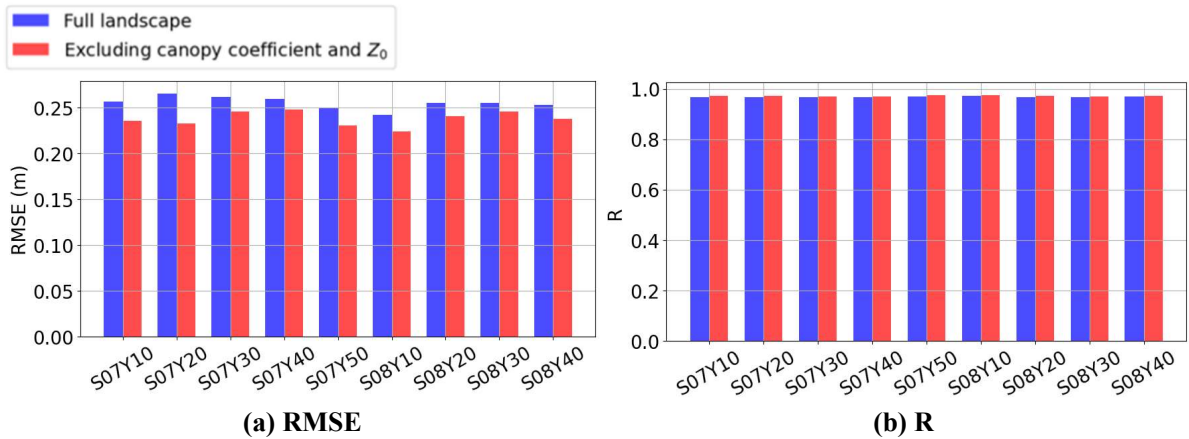

(a) RMSE  (b) R

Figure 8: Comparison of model performance including and excluding canopy and $Z_0$.

### 3.3. Storm set reduction

The general term of storm reduction or storm selection can refer to several distinct topics within the field of storm hazard assessment and modeling. Various optimal sampling methods—such as Bayesian quadrature (e.g., Toro et al. 2010), response surface methods (e.g., Resio et al. 2009), and genetic algorithms (e.g.,

Melby et al. 2021)—have been developed to select a small set of synthetic storms for numerical simulations. These methods aim to construct robust hazard curves while minimizing computational expense. This specific application is commonly referred to as storm selection for hazard curve construction or JPM-OS methods. Fischbach et al. (2016) systematically evaluated the performance of several storm reduction strategies within this context.

With advancements in ML-based surrogate model development, the cost of estimating storm responses has been significantly reduced. Consequently, it is now feasible to generate and utilize augmented storm suites containing tens of thousands of synthetic events for hazard curve construction (Nadal-Caraballo et al. 2022). This development has led to growing interest in storm selection for surrogate model training, which focuses on identifying the most informative storms in the surrogate model training process to optimize surrogate model performance. The storm reduction approach examined in this study specifically addresses this challenge. Identifying an optimal set of training storms can also be formulated as a space-filling problem (Liu et al. 2018; Viana 2013), where the goal is to evenly distribute the selected storms across the storm parameter space to capture the full range of information about the storm surge response function. Other interesting applications of storm selection might target identifying the most critical or high-impact storms for refined storm response simulations, often based on deaggregation analysis or potential impact-based methods (e.g., Liu et al. 2025a; b; Sohrabi et al. 2023). This is typically referred to as storm selection for refined modeling.

In this study, we first aimed to identify informative subsets from the full set of 645 synthetic storms. The goal was to select a subset that evenly covers the storm parameter space of all 645 storms, ensuring uniform representation across all possible storm characteristics. A $k$-medoids clustering algorithm (Schubert and Rousseeuw 2021) was employed to obtain representative storms (medoids) from the full dataset. Similar to $k$-means, $k$-medoids clustering partitions the data into $k$ clusters, but instead of computing an abstract centroid—which may not correspond to any real storm—it selects an existing storm as the cluster representative. The chosen medoid storm minimizes the average distance to all other storms in its cluster, making it interpretable as a physically meaningful representative. Although $k$-medoids clustering can be computationally intensive for large datasets, it is well-suited for selecting a "real" representative subset from the full 645 storm dataset. Both the S00Y00 and S09Y50 scenario datasets were used to evaluate the performance of storm subsets selected by the $k$-medoids clustering algorithm. In this analysis, storm parameters served as input features, and surrogate models were trained on the selected subsets and tested on the S00Y00 or S09Y50 scenario dataset with full 645 storms. The performance of subsets with the number of training storms ($n_s$) is presented in Figure 9, with the CMP2023 90 storms subset included for comparison.

The results demonstrate that as $n_s$ increases, both surrogate models trained on the S00Y00 and S09Y50 scenario dataset steadily improve their performance with respect to RMSE. Notably, the $k$-medoids algorithm selected a subset of 90 storms that outperforms the CMP2023 90 storms subset. Figure 10 presents a comparison of the distribution of the TC track and $P_c$ for the CMP2023 90 storms and $k$-medoids 90 storms. The visualization shows that the storms selected using the $k$-medoids clustering algorithm exhibit a more uniform and broader coverage of storm characteristics compared to the CMP2023 set. This suggests that the $k$-medoids algorithm is more effective in selecting storm subsets for surrogate model training. This finding aligns with the fact that the CMP2023 90 storm set was designed to minimize errors in hazard curve integration (Fischbach et al., 2021), rather than to optimize surrogate model performance at predicting surge for individual synthetic TCs. Nevertheless, the $k$-medoids algorithm effectively identifies representative cluster medoids that more uniformly capture the full coverage of the full 645 storm parameter space.

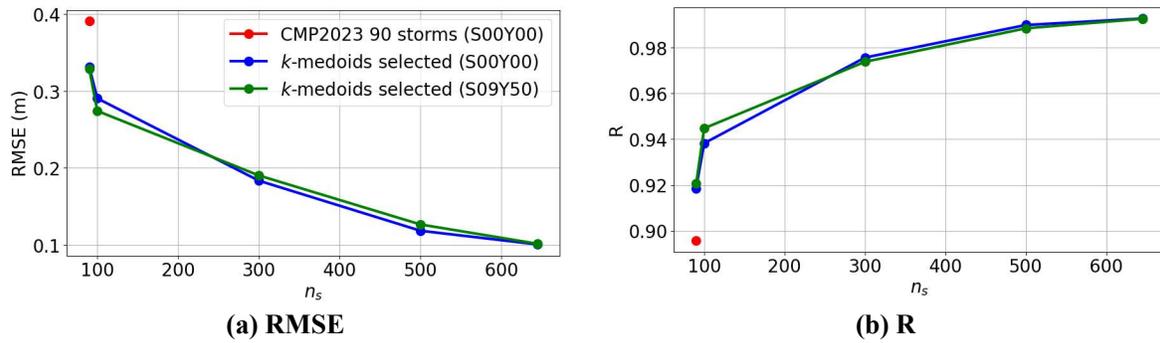

(a) RMSE  (b) R

**Figure 9: Performance of different size of training storm subset in S00Y00 and S09Y50 scenarios data.**

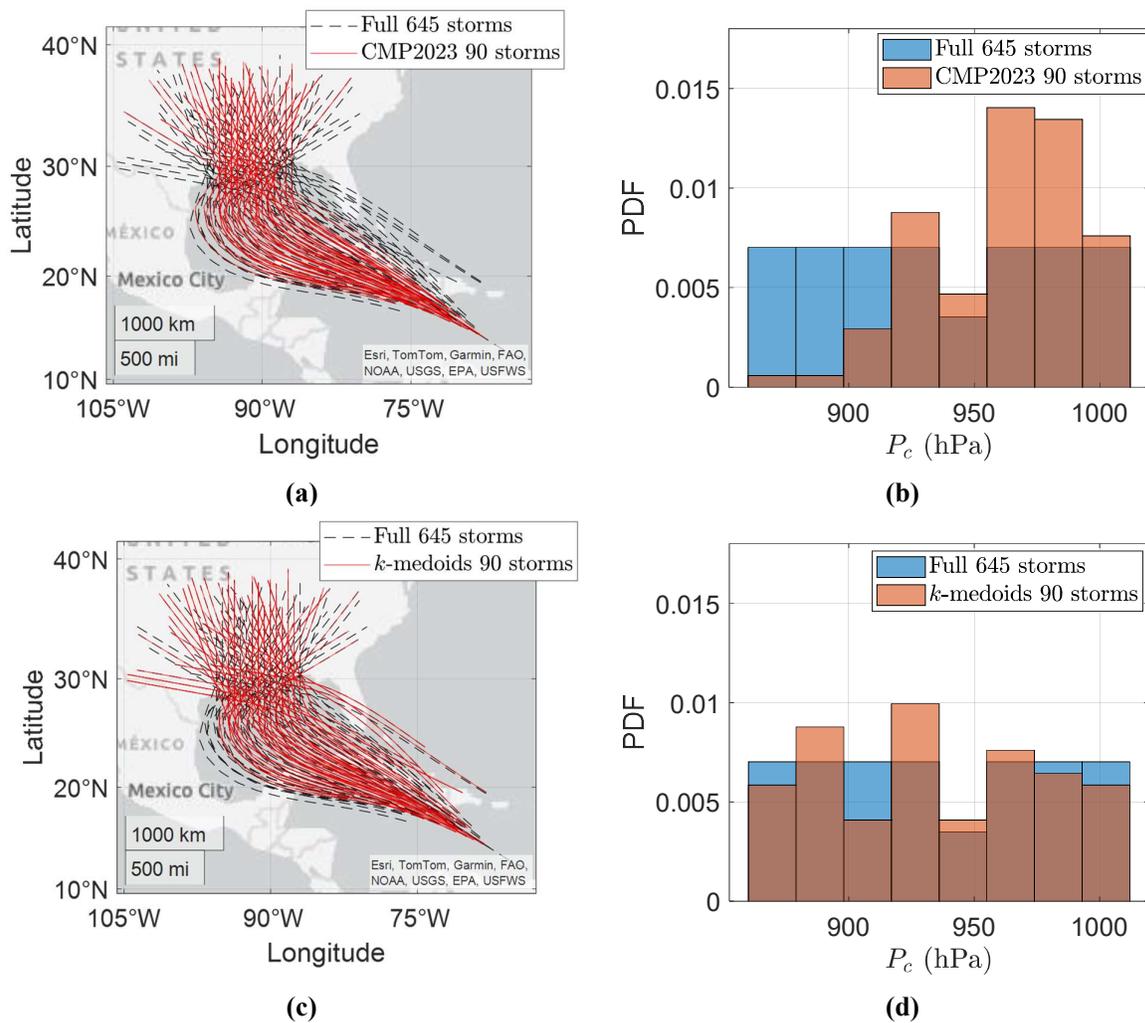

(a)  (b)

(c)  (d)

**Figure 10: Comparison of the distribution of the CMP2023 90 storms subset $k$-medoids 90 storms subset. (a) TC track distribution of CMP2023 90 storms subset; (b) $P_c$ distribution of CMP2023 90 storms subset; (c) TC track distribution of $k$-medoids 90 storms subset; (d) $P_c$ distribution of $k$-medoids 90 storms; Note in (b) and (d), histogram heights reflect normalized probability density function (PDF) value.**

In addition to the $k$-medoids clustering-based storm reduction, this study introduces a model performance–guided adaptive sampling algorithm. This adaptive algorithm is inspired by variance-based adaptive sampling strategies developed for kriging model design of experiments (Kyprioti et al. 2020; Liu et al. 2018). In the broader field of ML, adaptive sampling is also referred to as active learning, which is founded on the hypothesis that "if the learning algorithm is allowed to choose the data from which it learns—to be 'curious', if you will—it will perform better with less training" (Settles 2009).

The proposed algorithm in this work adaptively selects a subset of storms from the available synthetic storm suite to minimize training costs while maintaining reliable model performance. The design and implementation of this adaptive sampling procedure are illustrated in the algorithm flowchart shown in Figure 11. The key parameters are defined as follows: $n_{\text{int}}$ is the number of initial storms; s is the index subset of storms included for training; $n_s$ is number of storms included for training; $FRMSE_k$ is the RMSE of the trained surrogate model in the $k$th iteration averaged over all storms and GPs; $SRMSE_k^j$ is the RMSE of the trained surrogate model in the $k$th iteration averaged over all GPs but estimated for the $j$th storm of the whole storm suite separately; $\alpha$ is a threshold value set for target model performance; $s_k$ is the new storm index selected under the $k$th iteration.

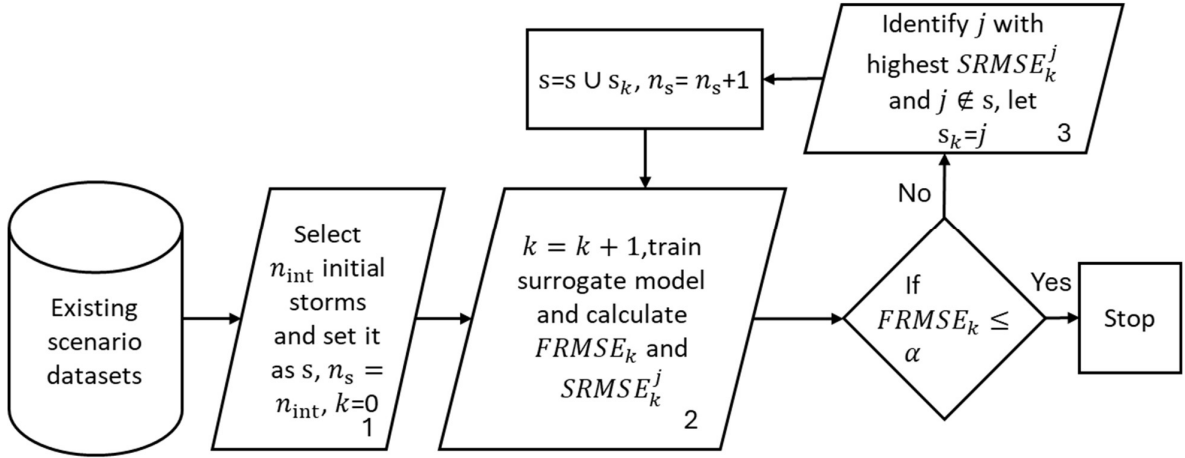

Figure 11: Storm adaptive sampling algorithm flowchart.

In Box 1, an initial storm subset is selected and used for the first round of model training. In Box 2, the surrogate model is trained, and the trained model's performance is evaluated both in terms of overall accuracy and on a per-storm basis. In Box 3, the storm associated with the maximum prediction error is identified and added to the training subset. This iterative process of training and testing is then repeated, progressively refining the model until the overall accuracy of the trained surrogate model reaches the target performance metric $\alpha$. Note that this adaptive sampling algorithm requires storm response data for all candidate storms and is more computationally expensive than the $k$-medoids clustering algorithm.

In this study, the adaptive sampling procedure was applied to our case study, taking into account the available data. Specifically, the goal is to select a smaller subset from the CMP2023 90 storm set, while the full set of 645 storms from the S00Y00 scenario dataset is always included in the model training process. To reduce the computational time in Box 2, $FRMSE$ and $SRMSE$ are calculated by a holdout cross-validation, where the model is trained on the S00Y00, S07Y10–Y50, and S08Y10–Y30 scenario datasets with reduced GPs, and tested on the S08Y40 scenario dataset. A leave-one-out cross-validation test, as conducted in Gharehtoragh and Johnson (2024), could serve as an alternative testing approach if additional computational resources are available or if higher accuracy is needed.

For the initial subset selection (Figure 11, Box 1), two approaches were tested: selecting storms using the $k$-medoids clustering method and selecting storms that produce the highest regional average storm surge. Re-sults indicate that the $k$-medoids clustering-based initial storm subset yields better overall model performance compared to the alternative approach. Various values of $n_{int}$ are tested, and the relationship between the number of storms ($n_s$) and model performance is presented in Figure 12. The results suggest that using $k$-medoids clustering to select the initial storm set, followed by adaptive sampling to iteratively expand it, provides an efficient and reliable strategy for storm selection in surrogate model training.

To further reduce the computational cost of this adaptive sampling method, an incremental learning approach has also been evaluated for the training procedure. In the step of Figure 11 Box 2, this approach updates the NN by fine-tuning it with newly selected storms rather than retraining the entire model. The results in this application return solid catastrophic forgetting issues (Chen and Liu 2018), where previously learned information was partially lost during incremental updates.

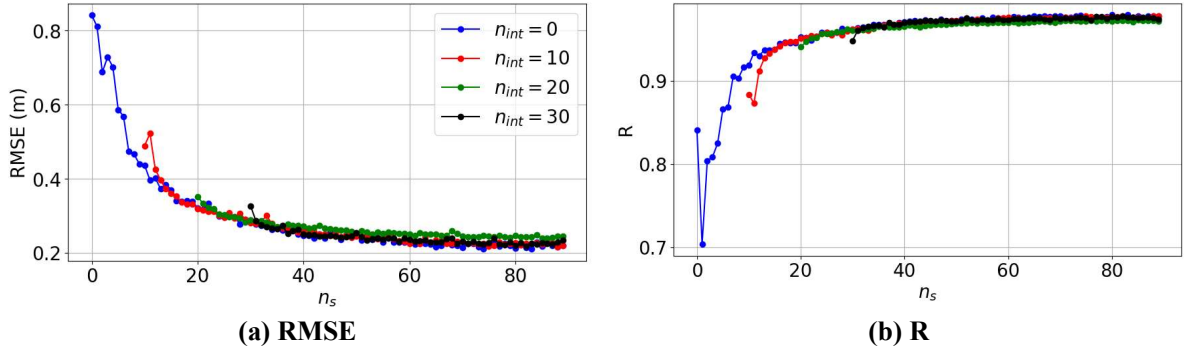

(a) RMSE  (b) R
Figure 12: Number of training storms vs. model performance in adaptive sampling.

### 3.4. Evaluation of reduction strategy

In this section, a new scenario dataset is used to evaluate the effectiveness of the reduction strategy developed in this study. Storm surge surrogate models are trained using all existing scenarios with a fixed GP subset containing 5,000 GPs, and reduced features—excluding canopy coefficient and $Z_0$ from the reference model input features, and reduced storm sets. The "unseen" S08Y50 scenario data is treated as the new scenario dataset and used to test the performance of the surrogate models trained with the reduction strategy.

To explore the effectiveness of the proposed strategy in a more general way, both an NN model and an XGBoost (Extreme Gradient Boost Tree) model are tested. The NN model adopts a similar architecture to the reference model (Gharehtoragh and Johnson 2024). The XGBoost model is included due to its widely recognized performance in general-purpose ML tasks (Chen and Guestrin 2016). The XGBoost library in Python is used, with model hyperparameters set as follows: maximum tree depth of 10, learning rate of 0.02, 9,000 boosting rounds, and early stopping with a patience of 10 rounds. The training storm set is evaluated by incrementally increasing the number of storms included from 10 to 90 in steps of 10. The storms are added in the order suggested by the adaptive sampling algorithm in the $n_{int} = 0$ case. Model performance across different storm set sizes is shown in Figure 13.

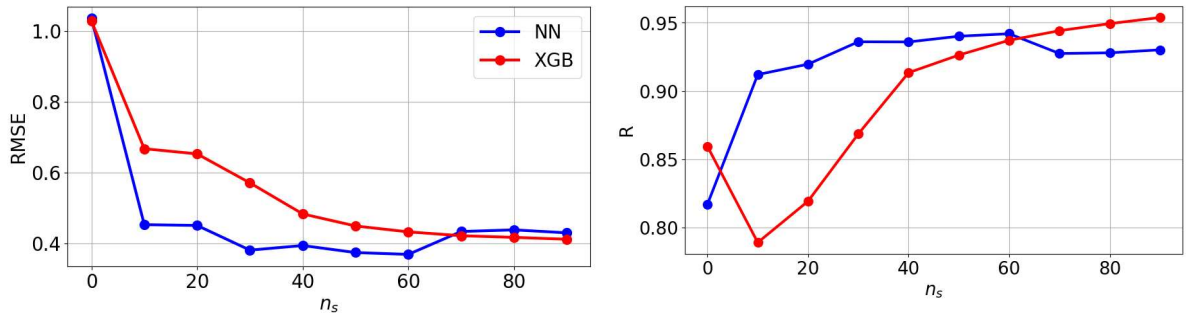

| (a) RMSE | (b) R |

**Figure 13: Number of training storms vs. model performance in evaluation.**

The results indicate that the NN model outperforms XGB models when $n_s$ is small. Both the NN and XGBoost models exhibit steadily improving performance as the number of training storms increases and converges to a maximum performance when $n_s$ reaches 90. This can be explained by recalling that the available testing data (i.e., the S08Y50 scenario's data) only contains the CMP2023 90 storms' storm surge response. It is noted that the RMSE in Figure 13 is higher than in Figure 12 because the storm surge values in the S08Y50 scenario dataset (testing data for Figure 13) are mostly higher than the S08Y40 scenario dataset (testing data for Figure 12). These findings demonstrate the efficacy of the proposed reduction strategy in developing surrogate models. For example, as indicated by Figure 13, when trained on 5,000 out of 80,000 grid points, using 10 out of 12 features and 60 out of 90 storms, the total size of training data downscale to around 5% of the full training data, but both NN and XGB model achieve performance of around R=0.94, which is comparable to that of models trained on the full dataset, while substantially reducing computational costs.

## 4. Storm selection for expanding training storms

In Sections 2 and 3, we introduced, applied, and evaluated the reduction strategy. In this section, we further investigate storm set expansion approaches. As noted in Section 2.2, the CMP 2023 dataset contains only 90 storms for the S07 and S08 scenario datasets, which limits its ability to represent the full space of TC parameters. With continued investment in CMP2029 and the availability of additional budget, it will become feasible to include more storms in the datasets. This raises a key question: How should we select additional storms to augment the existing 90 storms so that the resulting expanded storm set maximizes surrogate model training performance?

From a sampling-balance perspective, the expanded storm set should provide broad and representative coverage of the TC parameter space. At the same time, because 90 storms are already available, the new storms should not be overly similar to existing ones, as this could skew the training data distribution. Leveraging the storm selection frameworks discussed in this study, we compare several candidate methods for selecting additional storms, including clustering, farthest-point sampling (Gonzalez 1985; Hochbaum and Shmoys 1985), and adaptive sampling approaches. These storm expansion approaches are briefly introduced below.

(1) **CMP2023 + Farthest-Point Sampling (FPS):** In this approach, the CMP2023 90 storms are used as the initial subset. The farthest-point sampling algorithm is then applied to the remaining storms, iteratively selecting the storm farthest from the current subset and adding it to the expanded storm set.

(2) **CMP2023 + Clustered Farthest-Point Sampling (CFPS):** This approach combines $k$-means clustering with farthest-point sampling. First, the 645 storms are clustered into $k$ groups using k-means clustering. Within each cluster, farthest-point sampling is applied iteratively until the number of selected storms in each cluster reaches the target allocation $\lceil n_e/k \rceil$ where $n_e$ is the target number of storms in the expanded storm set, $\lceil \ \rceil$ represents the ceiling function. Note that this approach can not guarantee the number of storms in the expanded storm set is exactly $n_e$, because the initial storm set in some cluster might already exceed the target allocation $\lceil n_e/k \rceil$ and therefore result in a greater number of storms.

(3) **CMP2023 + Adaptive Sampling S00Y00 (AS_S00Y00):** The CMP2023 90 storms is used as the initial subset. The adaptive sampling algorithm introduced in Section 3.3 is then applied, using the S00Y00 scenario dataset for both training and testing, to iteratively add storms until the target number $n_e$ is reached. Note that this approach, as well as approach (4) requires storm surge response data to support the adaptive sampling algorithm.

(4) **CMP2023 + Adaptive Sampling S09Y50 (AS_S09Y50):** Similar to AS_S00Y00, except that the S09Y50 scenario dataset is used instead.

(5) **$k$-medoids Clustering:** The $k$-medoids clustering algorithm is applied directly to the full set of 645 storms, selecting $n_e$ cluster medoids as the representative storm subset. Note that this approach can not guarantee inclusion of the original CMP2023 90 storms in the expanded set.

Those approaches are comparatively tested using a target number $n_e = 140$ for storms in expanded storm sets. The performance of their trained surrogate models is evaluated by testing on the Y00S00 scenario dataset. A coefficient of variance for cluster proportion (CPCV) is computed to evaluate the uniformity of each resulting storm set. The CPCV is computed as:

$$CPCV = \frac{std(n_i^e)}{mean(n_i^e)} \quad (1)$$

where $n_i^e$ is the number of storms in the expanded storm set falls into the $i$th cluster; $i = 1 \ldots k$; the full 645 storm is clustered into $k$ cluster to calculate $std(n_i^{140})$ and $mean(n_i^{140})$; In this work, $k$ is set as eight;

Table 2 summarizes the information and performance of each storm set expansion approach. Relative to the CMP2023 90 storm set, all tested expansion approaches successfully reduced both the trained model error (RMSE) and the CPCV. Overall, the $k$-medoids clustering approach (selecting 140 storms directly without inclusion of the CMP2023 90 storms) and the adaptive sampling approaches (AS_S00Y00 and AS_S09Y50) produced similar and superior RMSE performance relative to CFPS and FPS. Notably, the $k$-medoids approach provided an additional benefit by achieving a substantially lower CPCV, indicating a more uniform representation of the TC parameter space. Although CFPS produced a higher RMSE than $k$-medoids and adaptive sampling, it achieved a lower CPCV than the adaptive sampling approaches. In contrast, FPS showed the poorest performance overall, exhibiting both the highest RMSE and the highest CPCV among all methods evaluated. Figure 14 plots the proportional distribution of each expanded storm set across the eight clusters. It reveals that the CMP2023 90 storm set exhibits a highly uneven cluster distribution. Notably, no storms fall into clusters 1 or 4, while clusters 6 and 7 contain disproportionately large proportions of storms. When the CFPS approach is applied, it partially enforces a more uniform distribution across clusters; however, the proportions in clusters 6 and 7 remain relatively high because the method is constrained by the original CMP2023 90 storm set distribution. The AS_S00Y00 and AS_S09Y50 approaches generate cluster distributions that are similar to each other.

It is important to note, however, that these results are influenced by the target size of the expanded storm set, $n_e$. As $n_e$ increases, the performance of CFPS is expected to improve, gradually approaching both the RMSE and CPCV achieved by the $k$-medoids clustering approach. Given that the $k$-medoids approach cannot guarantee the inclusion of existing storms in the expanded storm set, both the adaptive sampling and CFPS approaches may be more suitable, depending on $n_e$ and the distribution of existing storms.

Table 2: information and performance of storm set expansion approaches

| | Number of storms | If require surge response data | Trained model performance tested on the S00Y50 data set (RMSE, m) | Cluster proportion-coefficient of variance (CPCV) |
|---|---|---|---|---|
| Full 645 storms | 645 | | 0.10 | 0.24 |
| CMP2023 90 storms | 90 | | 0.39 | 1.16 |
| FPS | 140 | No | 0.27 | 0.67 |
| CFPS | 143* | No | 0.25 | 0.45 |
| AS_S00Y00 | 140 | Yes | 0.21 | 0.60 |
| AS_S09Y50 | 140 | Yes | 0.21 | 0.60 |
| $k$-medoids Clustering | 140 | No | 0.21 | 0.20 |

*Note: In the CFPS approach, the actual number of storms (143) is greater than $n_e$ (140), because the initial storm set (CMP2023 90 storms) in some cluster already exceeds target allocation $\lceil n_e/k \rceil$.

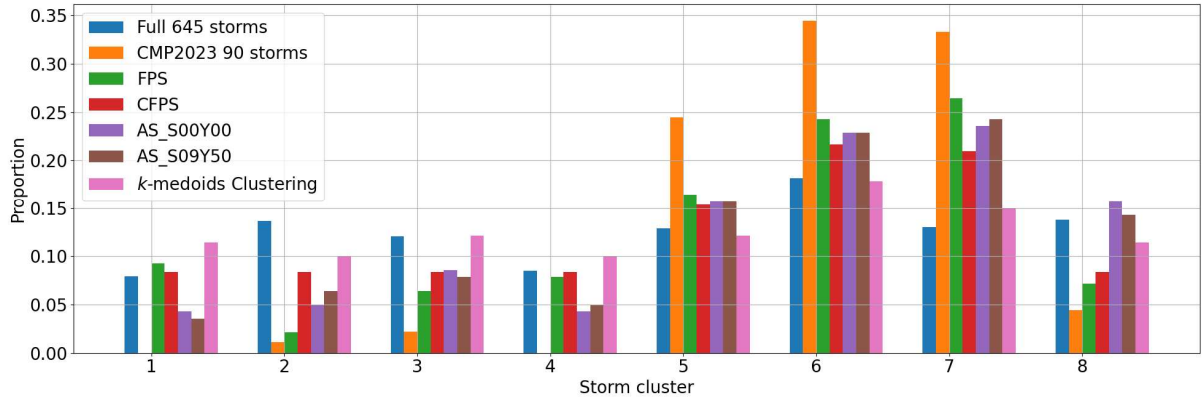

**Figure 14:** Cluster distribution of each expanded storm set.

## 5. Conclusion and discussion

This study developed a training data reduction strategy for storm surge surrogate model development. The resulting storm surge surrogate model is capable of predicting peak storm surge under varying landscape and climate scenarios. The proposed reduction strategy consists of three sequential steps: (1) Grid points (GPs) reduction; (2) Input features reduction; and (3) Storm set reduction. Among these steps, the GPs reduction (Step 1) incorporates the existing $k$-means clustering and Principal Components Analysis (PCA) methods (Kyprioti et al. 2021; Lee et al. 2021). The storm set reduction (Step 3) introduces a storm selection method designed to optimize surrogate model performance. This storm selection method is of particular interest because it also informs the reduction of synthetic storm numerical simulations, which are used to generate the surrogate model training data and represent one of the most computationally expensive components of storm hazard quantification.

The storm selection framework combines two complementary methods: $k$-medoids clustering and adaptive sampling. The $k$-medoids method is informed by the storm parameter space and aims to select a subset of storms that uniformly represent the full storm parameter distributions by clustering and identifying the centroid of each cluster as its representative storm. In contrast, the adaptive sampling approach is performance-driven—it iteratively identifies new storms to be added into the training datasets based on model errors and data needs, making it more computationally expensive but also more directly targeted for improving model accuracy.

The proposed reduction and training strategy was evaluated using various ML models. The results demonstrate that the approach is broadly applicable across different ML architectures, producing robust and efficient surrogate models. When trained on 5,000 out of 80,000 grid points, using 10 out of 12 features and 60 out of 90 storms, the total size of the training data is downscaled to around 5% of the full training data size. The trained model can achieve a performance of R = 0.94, which is comparable to that of models trained on the full dataset, while substantially reducing computational costs. Notably, the storm selection methods perform effectively across models, and the combined use of clustering and adaptive sampling provides flexibility in storm selection depending on data availability. This flexibility is particularly valuable in practical regional surrogate model development, where a clustering algorithm (lower cost) can be used to select an initial small storm suite for preliminary analyses, and additional storms identified by the higher-cost adaptive sampling algorithm can be incorporated later as computational resources permit and accuracy requirements increase.

It should be noted that the performance of the trained models in this study is constrained by the fact that most scenario datasets only contain storm surge simulations based on the Coastal Master Plan 2023 (CMP2023) 90 storm set. This set was originally selected to minimize errors in hazard curve integration prior to the conduction of this work. For the forthcoming Coastal Master Plan 2029 (CMP2029) simulations, it is recommended that a new storm subset be selected using the storm selection methodology proposed in this study to further improve surrogate model performance and generalization.

Leveraging the storm selection methodologies introduced in this study, we investigate several storm set expansion approaches for adding storms to an existing storm set—motivated by a practical case in which additional project budget becomes available to run more simulations. The results indicate that, among all tested methods, adaptive sampling provides the biggest improvement in trained surrogate model performance, whereas the Clustered Farthest-Point Sampling (CFPS) approach yields an expanded storm set with more uniform coverage of the TC parameter space.

## 6. Data availability

Data for this project are available at DesignSafe-CI, https://doi.org/10.17603/ds2-0ksb-yy40.

## 7. Acknowledgements

This work was supported by the U.S. National Science Foundation under awards 2238060 and 2118329. We thank members of the modeling teams for Louisiana's Comprehensive Master Plan for a Sustainable Coast and funding from the Coastal Protection and Restoration Authority, under the 2023 and 2029 Coastal Master Plan's Master Services Agreement, for the original simulation data used in this study. Any opinions, findings, conclusions, and recommendations expressed in this material are those of the authors and do not necessarily reflect the views of the funding entities.

**Competing Interests**

The authors have no relevant financial or non-financial interests to disclose.

**Author Contributions**

ZL: Conceptualization, Methodology, Software, Investigation, Formal analysis, Data Curation, Visualization, Writing—Original Draft, Writing—Review & Editing. MAG: Conceptualization, Methodology, Software, Data Curation, Writing—Review & Editing. BKL: Conceptualization, Writing—Review & Editing. DRJ: Conceptualization, Methodology, Data Curation, Writing—Review & Editing, Supervision, Funding Acquisition, Project administration.